\documentclass[aps,prf,showpacs,floatfix,longbibliography]{revtex4-1}
\usepackage{amsfonts}
\usepackage{natbib}
\usepackage{amsmath}
\usepackage{amsthm}
\usepackage{graphics}
\usepackage{graphicx}
\usepackage{subfigure}
\usepackage{amssymb}
\usepackage{xcolor}
\usepackage{soul}
\usepackage[T1]{fontenc}
\usepackage{amsmath,amssymb,color,epsfig,latexsym,natbib,graphicx,dcolumn}

\newcommand{\ou}{\bf \overline{u}}
\newcommand{\bu}{{\bf u}}
\newcommand{\tu}{\tilde{\bf u}}

\def\NEW#1{{\textcolor{black}{#1}}}
\def\RMV#1{{\textcolor{gray}{  }}}
\begin{document}

\title{ 
Symmetry breaking in a turbulent environment 
}

\author{Alexandros Alexakis}
\affiliation{Laboratoire de Physique de l'\'Ecole Normale Sup\'erieure, CNRS,
    PSL Research University, Sorbonne Universit\'e, Universit\'e de Paris,
    F-75005 Paris, France}

\author{Santiago J. Benavides}
\affiliation{Department of Earth, Atmospheric, and Planetary Sciences, Massachusetts Institute of Technology, Cambridge, MA 02139, USA}

\author{Kannabiran Seshasayanan}
\affiliation{Service de Physique de l'Etat Condens{\'e}, CNRS UMR 3680, CEA Saclay, 91191 Gif-sur-Yvette, France}
\affiliation{Department of Physics, Indian Institute of Technology Kharagpur, Kharagpur 721 302, India}

\author{Fran\c{c}ois P\'etr\'elis}
\affiliation{Laboratoire de Physique de l'\'Ecole Normale Sup\'erieure, CNRS,
    PSL Research University, Sorbonne Universit\'e, Universit\'e de Paris,
    F-75005 Paris, France}

\date{\today}

\begin{abstract} 
 
In this work we investigate symmetry breaking in the presence of a turbulent environment. The transition from a symmetric state to a symmetry-breaking state is demonstrated using two examples: (i) the transition of a two-dimensional flow to a three dimensional flow as the fluid layer thickness is varied and (ii) the dynamo instability in a thin layer flow as the magnetic Reynolds number is varied. We show that these examples have similar critical exponents that differ from the mean-field predictions. The critical behavior can be related to the multiplicative nature of the fluctuations and can be predicted in certain limits using results from the statistical properties of random interfaces. 
Our results indicate the possibility of existence of a new class of out-of-equilibrium phase transition controlled by the multiplicative noise.

\end {abstract}
\maketitle

Phase transitions are ubiquitous in nature. The liquid-gas transition or the transition from a magnetized to a non-magnetized state in ferromagnetic materials are textbook examples  \cite{Ma,goldenfeld,zinn2002quantum}. Critical phenomena of continuous phase transitions have  been a major research topic for more than 50 years. It is now well understood that, at equilibrium, the thermal fluctuations play a dominant role: the amplitude  of the order parameter, say $A$, depends on the distance from the critical point, say $\mu$, as a power-law $A\propto \mu^\beta$ where the value of the exponent $\beta$ differs from the mean-field prediction obtained when thermal fluctuations are neglected. These results are verified in experiments and are understood theoretically for instance through renormalization methods. In contrast, the behavior  of critical phenomena in  non equilibrium systems remains \NEW{ less  well understood}. In liquid crystals, a transition between two topologically different nematic phases was shown to belong to the class of directed percolation \cite{Kaz2007}. The transition from the laminar state to turbulence in extended shear flows \cite{lemoult2016directed, sano2016universal, chantry2017universal} is an out of equilibrium phase transition that  also belongs to the directed percolation universality class.
\begin{figure}                                       
  \centering
  \includegraphics[width=0.45\textwidth]{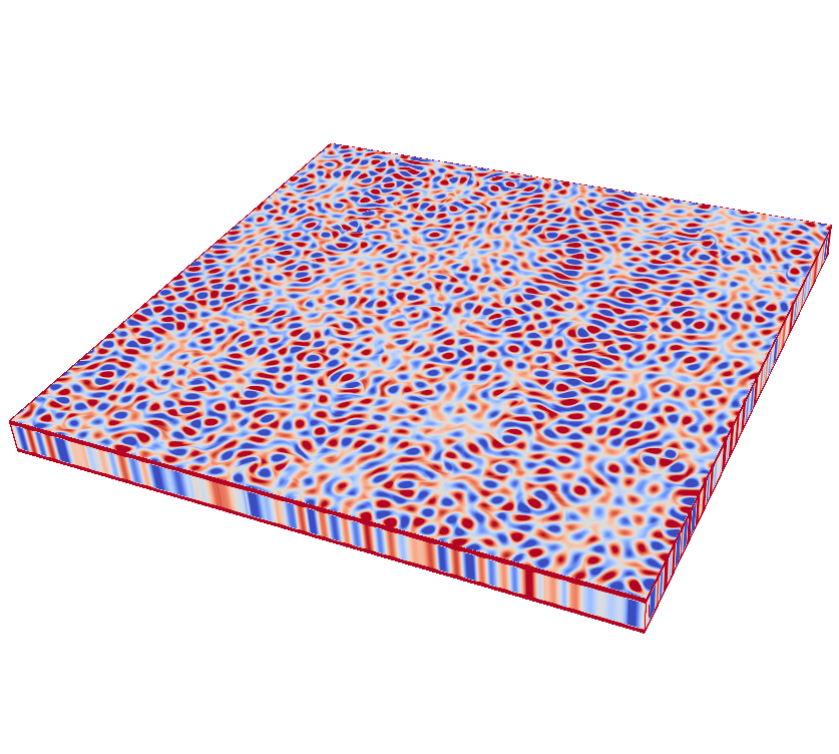}
  \includegraphics[width=0.45\textwidth]{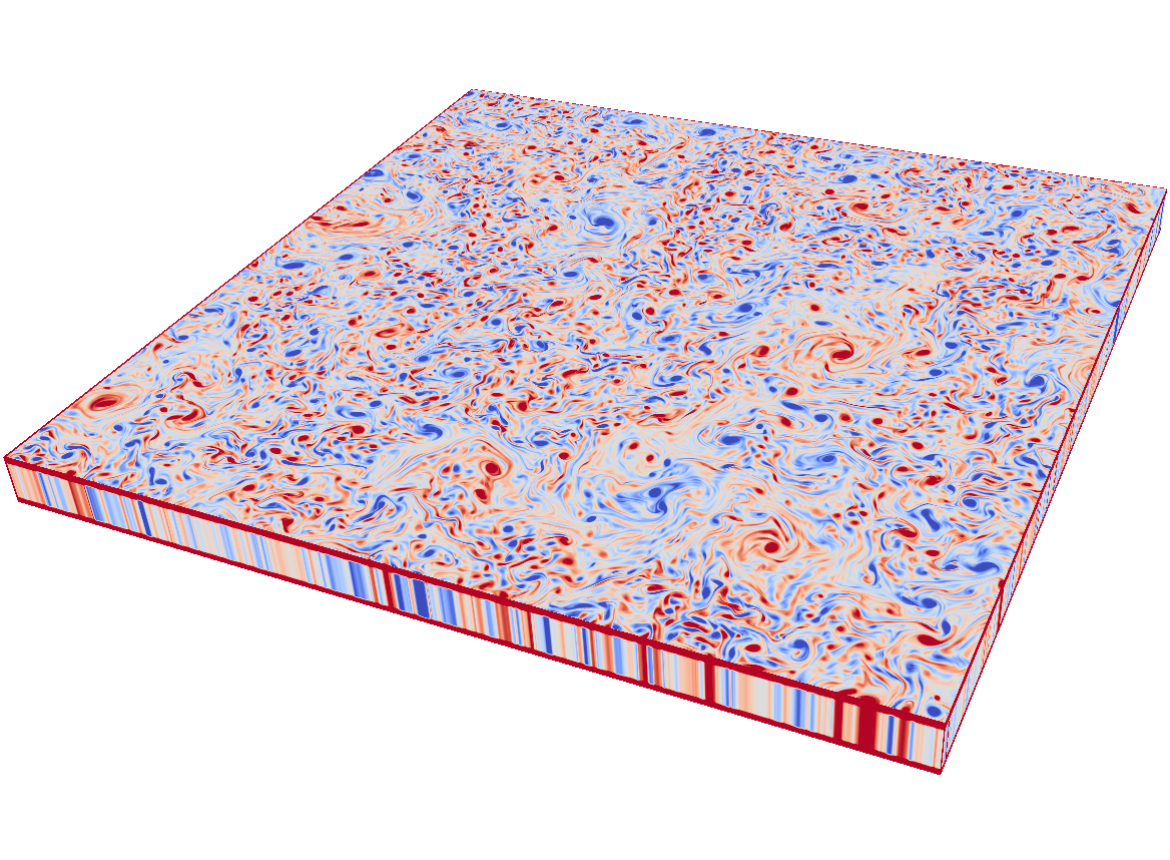}
  \caption{Vertical vorticity, $\omega_z = \hat{z}\cdot\left(\nabla \times {\bf \overline{u}}\right)$, of the 2D field for a random (left) and a turbulent state (right).} \label{2Dflow1}
\end{figure}                                          
Here we consider examples of bifurcations over a turbulent flow, in which the system transitions from a state that respects a certain  symmetry to a different state where this symmetry is broken. 
\NEW{We need to emphasise, that the transition is from a turbulent/chaotic state to an other turbulent state and thus it differs from the classical laminar to turbulent transition. Furthermore,}
the symmetries and the  nature of the coupling of the turbulent fluctuations differ from the former examples indicating the possibility of a new universality class.

The first system that we consider is a two dimensional (2D) flow which undergoes an instability towards a three dimensional (3D) flow.  The nature of the transition from a 2D to a 3D flow is a challenging  topic of wide-range interest in turbulence \cite{alexakis2018cascades}. It is a common situation in geophysics as rotation and the small pressure scale height of planetary atmospheres tend to bidimensionalize the flows \cite{byrne2013height,young2017forward}. Here we consider an idealized flow confined in a thin layer of thickness $H$ in the normal $z$-direction and of width $L\gg H$  in the in-plane $x$ and $y$ directions with free slip boundary conditions in  $z$ and periodic boundary conditions in  $x$ and $y$. The flow is described by the incompressible velocity $\bf u$ that follows the Navier-Stokes equations,
\begin{equation}
    \partial_t {\bf u + u \cdot \nabla u = -\nabla} P + {\bf \nu \nabla^2 u -\alpha \overline{u} +f},
\end{equation}

Where $P$ is the pressure, $\nu$ is the kinematic viscosity and $\alpha$ is a drag coefficient that acts only on the vertically averaged part of the flow, denoted by $\overline{\bf u}$, used to model Ekman friction \cite{pedlosky}. Energy is injected by ${\bf f}$ a random delta-correlated in time forcing, with a fixed averaged energy injection rate $\epsilon$\NEW{, an input parameter}.
\NEW{It is two-dimensional, depending only on $x$ and $y$, so that $\overline{\bf f}={\bf f}$, and acts only on the horizontal components. It is 
acting at some length scale $\ell$, such that $H \ll \ell \ll L$}. 
\NEW{The injection rate $\epsilon$ and the length scale $\ell$ will be used to nondimensionalize our system and will be set accordingly to unity. }
The part of the flow that varies along the vertical direction is denoted as $\tilde{\bf u} = {\bf u -\overline{u}}$ and follows the equation
\begin{equation}
    \partial_t \tilde{\bf u} 
 + \overline{\bf u} \cdot \nabla \tilde{\bf u} 
 + \tilde{\bf u} \cdot \nabla \overline{\bf u}= 
  \overline{\tilde{\bf u} \cdot \nabla \tilde{\bf u}}
 - \tilde{\bf u} \cdot \nabla \tilde{\bf u}
 - {\bf \nabla} \tilde{P} 
 + \nu {\bf \nabla^2 \tilde{u}}.
\end{equation}
\NEW{Note that because $\overline{\bf f}={\bf f}$ the velocity variation $\tilde{\bf u}$ is not directly forced, so that $\tilde{\bf u}=0$ is always a solution
of the system. }
For very thin layers and close to the onset of the instability $\tilde{\bf u}$ can be approximated with one Fourier mode in the $z$-direction as in \cite{benavides2017critical}. 

For small $H$ a purely two dimensional  flow is generated, for which the velocity field is planar and invariant under translation across the layer. Its dynamics is  determined by the value of 
the Reynolds numbers $Re={\epsilon^{1/3}\ell^{4/3}}/{\nu}$ and 
$R_\alpha={\epsilon^{1/3}}/{\alpha\ell^{2/3}} $. For small $Re,R_\alpha$ the flow is random following the statistical properties of the forcing with Gaussian fluctuations, and a limited range of length-scales excited. In contrast, for large values of $Re$ and $R_\alpha$ the flow is turbulent and a cascade develops leading to fluctuations with non-Gaussian statistics distributed over a wide range of scales. We will refer to these limiting cases as {\it random} and {\it turbulent} respectively.  
A snapshot of the vertical vorticity, $\omega_z = \hat{z}\cdot\left(\nabla \times {\bf \overline{u}}\right)$, is displayed in Fig. \ref{2Dflow1} for a random (left panel) and a turbulent (right panel) state. 
In both cases, the fluctuations do not depend on the vertical coordinate and the system is invariant in this direction $\bf \tilde{u}=0$. If however $H$ is increased, the flow breaks this symmetry and three-dimensional variations \NEW{become unstable} $ \tilde{\bf u}\ne0$.
The system thus changes from a phase where $\bf \tilde{u}=0$ pointwise to a phase where $ \tilde{\bf u}\ne0$
at a critical height that is shown to scale like $H\propto \ell Re^{-1/2}$ \cite{benavides2017critical}. In this system we use as control parameter $\mu$ the normalized height of the layer $\mu=H/\ell$ while the order parameter is characterized by the different moments of the 3D fluctuations $A_m = \langle |\tilde{\bf u}|^m\rangle $ where the angular brackets stand for space-time-averaging. The size of the system is measured by the parameter $\Lambda=L/\ell$.


The second system that  we  investigate is the dynamo instability of a swirling electrically conducting fluid transitioning from an unmagnetized to a magnetized state \cite{moffatt2019self}. The system is governed by the equations of magneto-hydrodynamics (MHD)
\begin{align}
{\partial_t 
\bu + \bu \cdot \nabla \bu} & = -\nabla P + \nu \nabla^2 \bu -\alpha \ou + {\bf b\cdot \nabla b+f} \\
\partial_t {\bf  b + \bu \cdot \nabla b} & = {\bf b\cdot \nabla \bu}  + {\bf \eta \nabla^2 b }
\end{align}
where $\bf b$ is the magnetic field and $\eta$ the magnetic diffusivity. As in the previous case  the considered flow is confined in a thin layer, here with triple periodic boundary conditions. It is important to note that in the absence of the third component no dynamo instability exists. For this reason, although the forcing is invariant along the $z$-direction as before, all three components are present \NEW{in $\bf f$ for this problem }({\it i.e.} \NEW{two-dimensional, three-component,} 2D3C). It is again random and injects energy at a typical length $\ell$ at rate $\epsilon$. The ratio of the energy injection rate in the transverse component $\epsilon_v$ to the energy injection rate in the in-plane directions $\epsilon_h$ is  measured by  $\gamma = \epsilon_v/\epsilon_h$. The layer thickness $H$ is sufficiently thin so that the flow remains 2D3C $\bu=\ou$ while it is thick enough so that a single Fourier mode of the magnetic field becomes unstable  ${\bf b}(x,y,z,t)={\tilde{\bf b}(x,y,t)e^{i2\pi z/H}}$ as in \cite{seshasayanan2018growth, seshasayanan2017transition}. Keeping, $Re,R_\alpha,\Lambda$ (defined as before) fixed we use the magnetic Reynolds number $\mu = Rm = \epsilon_h^{1/3} \ell^{4/3} / \eta$ as control parameter and as order parameter the different moments of the magnetic field 
$A_m = \langle |\tilde{\bf b}|^m \rangle$.  
\begin{figure}
  \centering
  \includegraphics[width=0.45\textwidth]{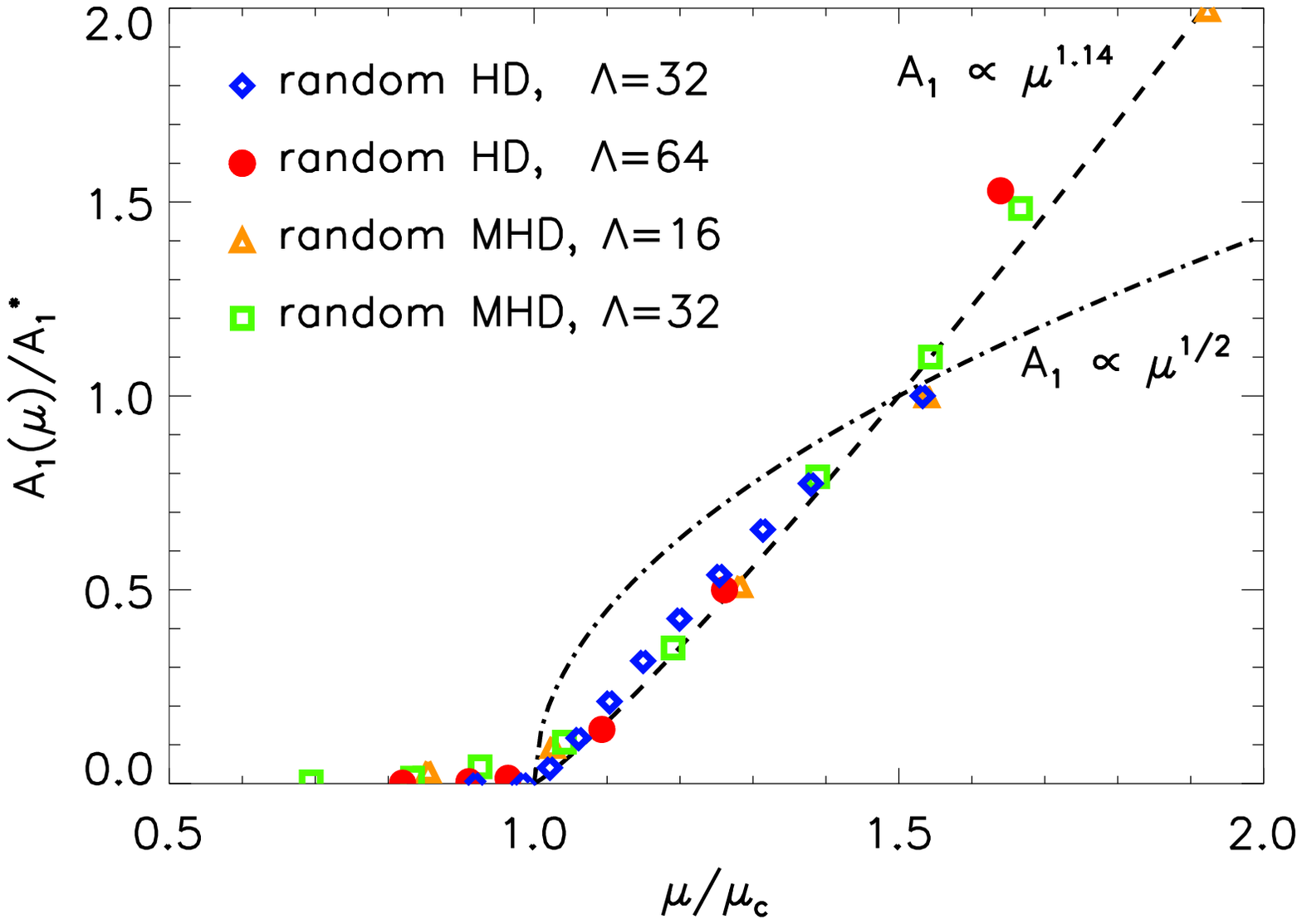}
  \includegraphics[width=0.45\textwidth]{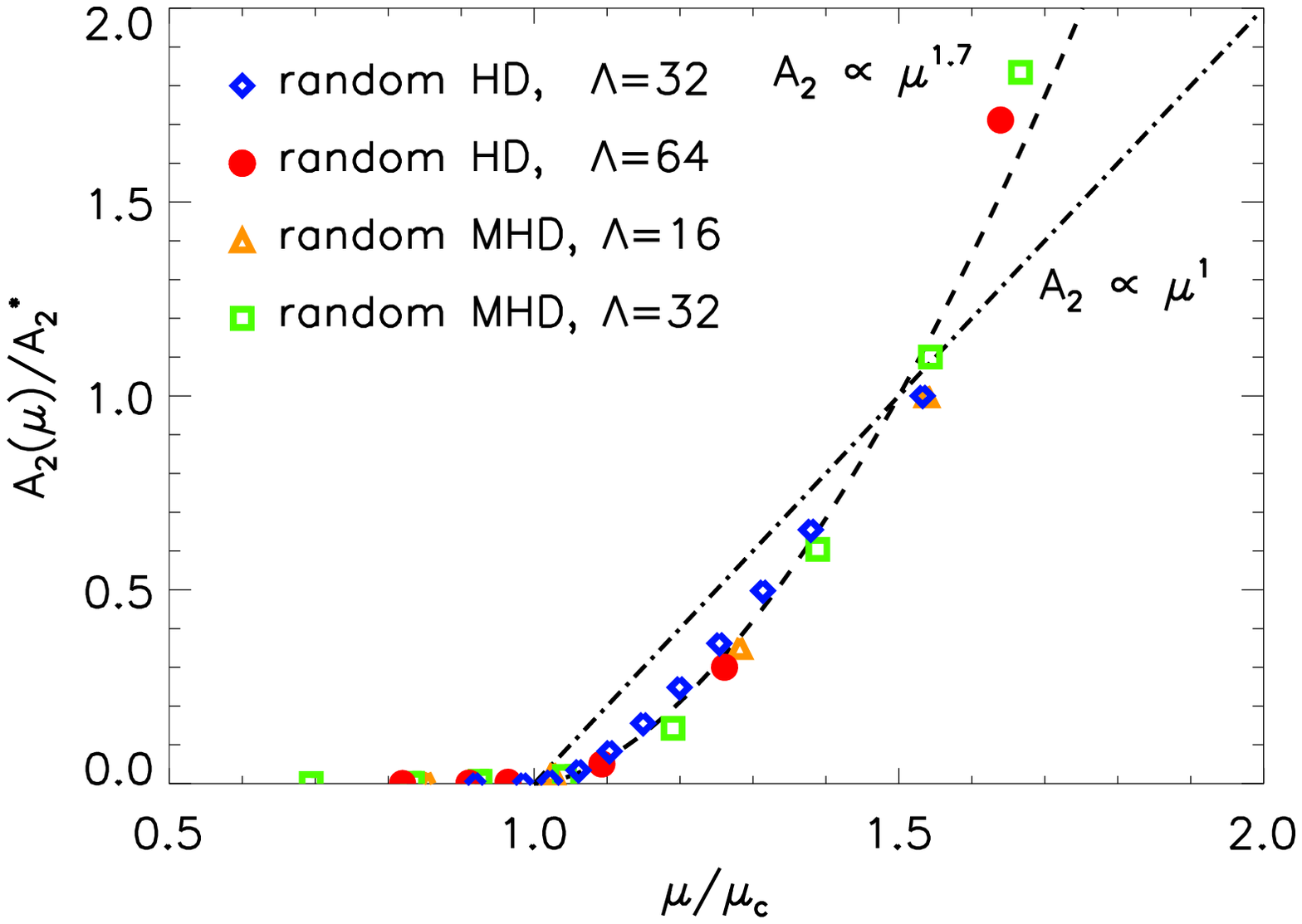}
  \caption{First and second moment $A_1,A_2$ for both the hydrodynamic problem and the MHD problem and for a random flow and different values of $\Lambda$.  The $y$-axis is normalised by $A_m^*=A_m(3\mu_c/2)$. }\label{fig:2Dgrowth2}
\end{figure}                                                                        
The two systems are numerically simulated using codes described in \cite{benavides2017critical,seshasayanan2017transition} on a $2048^2$ grid. The simulations were run until a statistically steady state is reached in which the different moments are measured.

We begin by examining the random flow, for which $Re\simeq R_\alpha \simeq1$, depicted in the left panel of Fig. \ref{2Dflow1}. 
The amplitudes $A_1$ and $A_2$ are displayed in Fig. \ref{fig:2Dgrowth2} as a function of $\mu$ for the hydrodynamic model (HD) and the MHD model with $\gamma=4$, for different values of $\Lambda$.
For both systems the first and the second moment collapse on a single master curve. 
\NEW{ Independence of the data on $\Lambda$} also indicates that the large box limit has been reached. 
As a consequence both systems appear to have the same critical behavior, suggesting a possibility that they belong to the same universality class.
 The moments bifurcate from zero at a critical value $\mu=\mu_c$
and scale with $\mu-\mu_c$ as power laws: $A_1\propto (\mu-\mu_c)^{\beta_1}$ and
$A_2\propto (\mu-\mu_c)^{\beta_2}$.
An accurate estimate of the value of the exponents $\beta_1,\beta_2$ is difficult to obtain. For these turbulent systems, the existence of low frequency velocity fluctuations renders the situation difficult as statistical convergence requires very long simulations. 
\NEW{ However, one can say with confidence that they clearly differ from $\beta_1=1/2$ and $\beta_2=1$
that are the exponents obtained for static fields or by mean-field predictions where the small scale fluctuations
are modeled by tranport coefficients like an eddy diffusivity or an alpha coefficient \cite{alexakis2018effect}. }
They also differ from the zero dimensional $d=0$ bifurcations in the presence of multiplicative noise that is termed  on-off intermittency and leads to $\beta_1=\beta_2=1$ \cite{fujisaka1985new,platt1993off}.

\begin{figure*}
  (a)\includegraphics[width=0.95\textwidth]{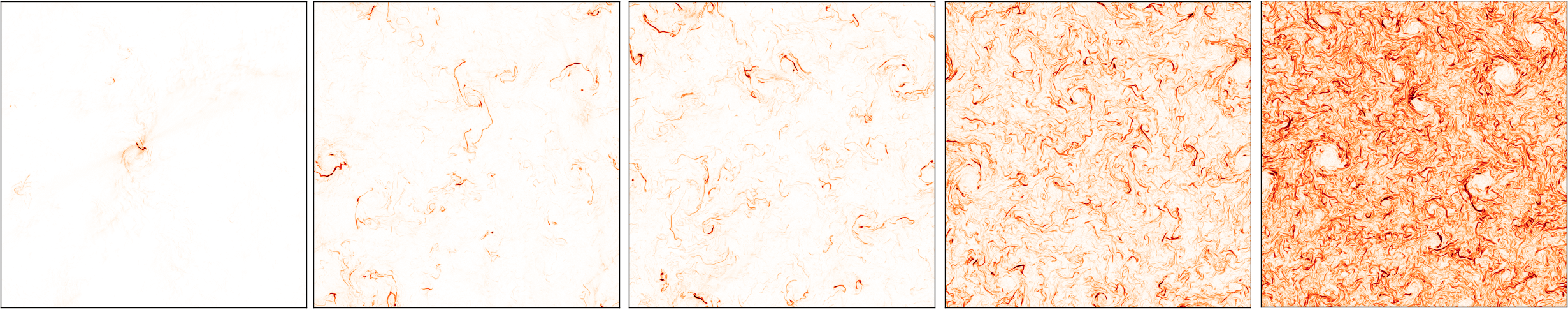}
  (b)\includegraphics[width=0.95\textwidth]{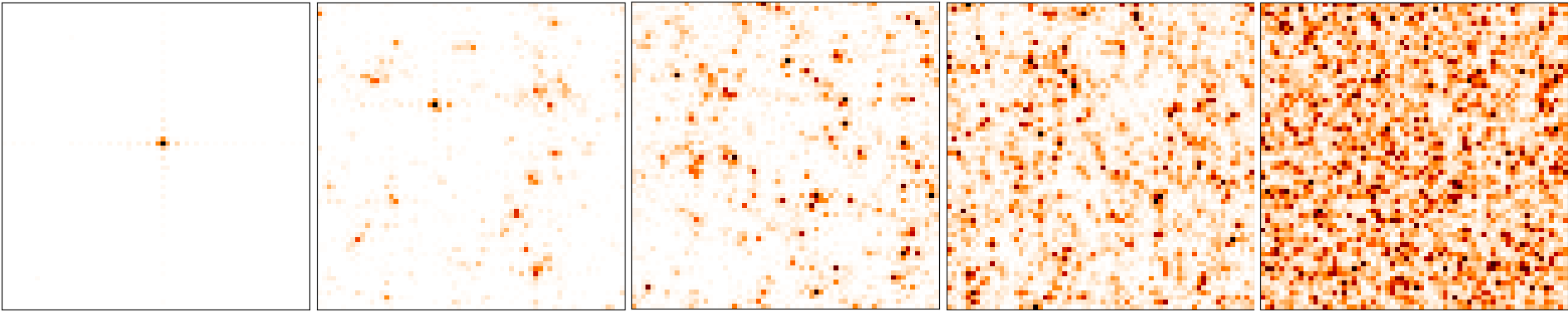}
  (c)\includegraphics[width=0.95\textwidth]{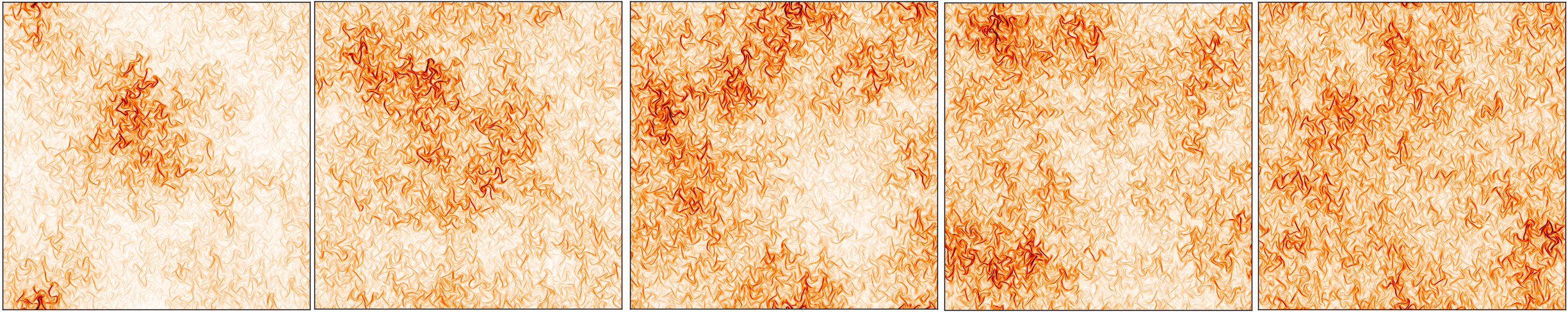}
  \caption{
   For increasing values of $\mu$ (from left to right and starting from close to $\mu_c$)    the images display
  (a) Energy density of 3D velocity field for the random base flow for the data points displayed in Fig. \ref{fig:2Dgrowth2}.
  (b) Energy density of the field $\phi$ for the field equation, Eq. (\ref{eqfield}) with white noise solved on a $64\times64$ grid.
  (c) Energy density of magnetic field energy for the data points displayed in Fig. \ref{fig:2Dgrowth4}.
      }\label{fig:fields}
\end{figure*}                           
In order to explain these new exponents and to identify and characterize the universality class of these systems, we resort to \NEW{deriving} a field equation\NEW{, modeling the approximate equations of motion for the amplitude of the unstable $\tilde{\bf u}$ and ${\bf b}$ near the threshold of instability. This derivation will be based on symmetries of the bifurcating system}. 
%
%

\NEW{The hydrodynamic problem is symmetric under reflection in the $z=0$ plane, which we denote  by $\cal S$.  Once $\mu$ goes beyond the critical value, the first linearly unstable vertical mode breaks this planar symmetry and is thus odd under $\cal S$. If we  denote this solution $\tilde{\bf u} = \phi(x,y,t){\bf v_u}$, where $\phi$ is the amplitude of the unstable mode and ${\bf v_u}$ is the vertical mode structure, then we have that $\cal S {\bf v_u}= - {\bf v_u}$. Because the hydrodynamic problem is symmetric under reflection,  if $\phi {\bf v_u}$ is a solution then $\cal S \phi {\bf v_u}=-\phi {\bf v_u} $ is also a solution. In other words $\phi$ and $-\phi$ are solutions of the problem.}
%
\NEW{Similarly, for the magnetic problem, because of the invariance of the MHD equation under change of sign of the magnetic field,  if  $ {\bf b}$ is a solution so is $-{\bf b}$. With the same reasoning let ${\bf b_u}$ be the linearly unstable mode and $\phi$ its amplitude, if $\phi {\bf b_u}$ is a solution, so is $-\phi {\bf b_u}$. }
\NEW{
It is important to notice that these symmetries are satisfied even taking into account  the turbulent fluctuations. Therefore when modelling  the effect of the turbulent fluctuations in the field equation by stochastic terms,  only odd terms in $\phi$ appear.
Accordingly, the first order term in $\phi$ that couples to the spatio-temporal fluctuations of the background field is linear. 
The symmetries of the problem thus imply that the noise acting on the perturbation field is multiplicative. For the same reason, the lowest order  nonlinear term is cubic. }

We thus end with the following field equation 
\begin{equation}
\frac{\partial \phi}{\partial t}=\mu  \phi - C \phi^3+ D\nabla^2 \phi +\zeta({\bf x},t) \phi
\label{eqfield}
\end{equation}
where $\zeta$ is spatio-temporal noise (interpreted in the Stratonovich sense), $\mu$ is the control parameter and $C$, $D$ are constants. 
\NEW{Here, the term $\zeta({\bf x},t) \phi$ expresses the local amplification or decrease effects, while $\mu \phi$ expresses their mean counter-parts.
The non-linearity $- C \phi^3$ is responsible for saturating the growth. 
The term $D\nabla^2 \phi$ is responsible for diffusing any localized structure of $\phi$.}
This equation has been studied to model for instance chemical reactions or synchronization transition \cite{munoz,tauber2014critical,henkel2008non}. 

 When $\zeta$ is white and Gaussian, renormalization group methods allow to predict the critical behavior of the system.
 For a space of dimension $d\le 2$, a transition exists between an absorbing  phase where $\phi=0$ and an active phase where $\phi\ne0$. Close to the critical point, the field scales as  $\langle |\phi|^n \rangle=(\mu-\mu_c)^{\beta_n}$. 
It has been shown \cite{Tu} that some critical exponents of Eq. (\ref{eqfield}) can be related to the exponents of the Kardar-Parisi-Zhang equation (KPZ) \cite{KPZ,halpin}. Indeed, the linear part of Eq. (\ref{eqfield}) is transformed by the Cole-Hopf transformation into the KPZ equation. This equation describes the growth of a random surface when nonlinear effects are taken into account. Some predictions of the KPZ equation are thus useful for the systems that we are considering. For $d=2$ and white noise, the exponents $\beta_n$ have been calculated numerically  $\beta_1\simeq1.14$ and $\beta_2\simeq1.7$ \cite{genovese}. These predictions are displayed in Fig. \ref{fig:2Dgrowth2} and are compatible with the results obtained for the two systems under study. 

The behavior of the different moments $A_m$ results from  the spatial distribution of the unstable field ($\tu, {\bf b}, \phi$). In the top panels (a) of Fig. \ref{fig:fields} five snapshots of the energy density of the field $\tu$ are shown for  different values of $\mu$.  The snapshots correspond to the data marked by blue diamonds in Fig. \ref{fig:2Dgrowth2}. Far from the onset (rightmost panels)  the unstable field is spread throughout the domain. As $\mu$ comes closer to the onset the unstable field becomes more sparse occupying a smaller and smaller fraction of the domain. Very close to the onset (leftmost panel) only a few structures are left and in most of the domain the unstable field is almost zero. In panel (b) a  series is shown for solutions of Eq. (\ref{eqfield}) that shows similar features.

There are a few remarks that need to be made here. First we stress that the predictions for the field equation, Eq. (\ref{eqfield}), hold for the limit of infinite domain size $\Lambda\to\infty$. For finite domains these exponents can be contaminated by finite size effects \cite{barber1983finite}. One can see for example from Eq. (\ref{eqfield}) that when the inverse diffusion time scale $L^2/D$ is much smaller than the growth rate fluctuations, the spatial fluctuations are averaged out and the system recovers the mean field behavior. This limitation has a profound implications on the systems under study because the domain size is always finite and diffusion is controlled by eddy-diffusion that in general has non-trivial dependency with the system control parameters. For example, in the MHD system when we decrease the parameter $\gamma$ we decrease the growth-rate that depends on the product of vertical and horizontal velocity components while we increase the turbulent diffusivity that depends only on the horizontal components. As a result the system becomes much more diffusive as $\gamma$ is decreased.
\begin{figure}[!ht]                                                                          
\centering
\includegraphics[width=0.45\textwidth]{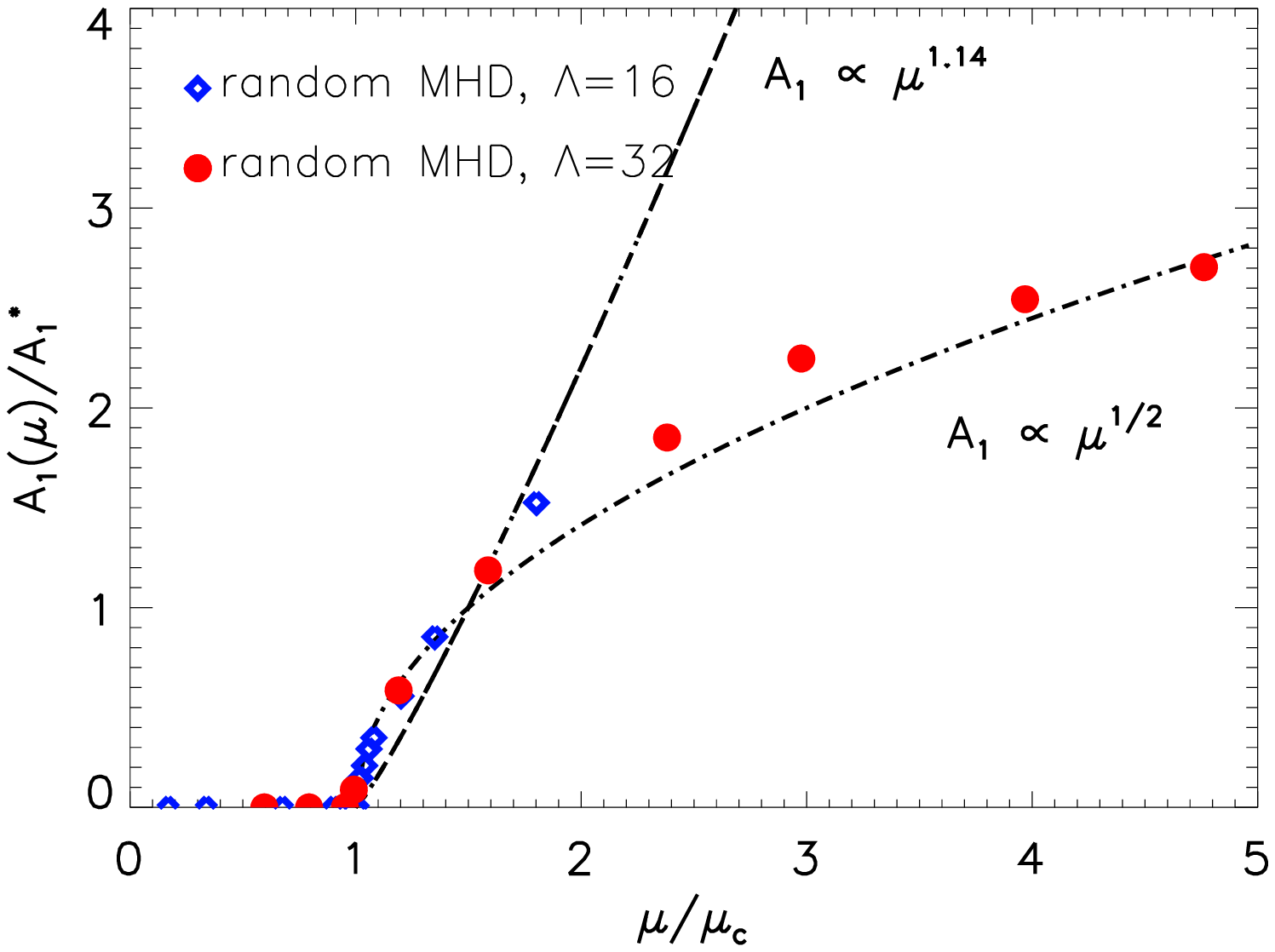}
\includegraphics[width=0.45\textwidth]{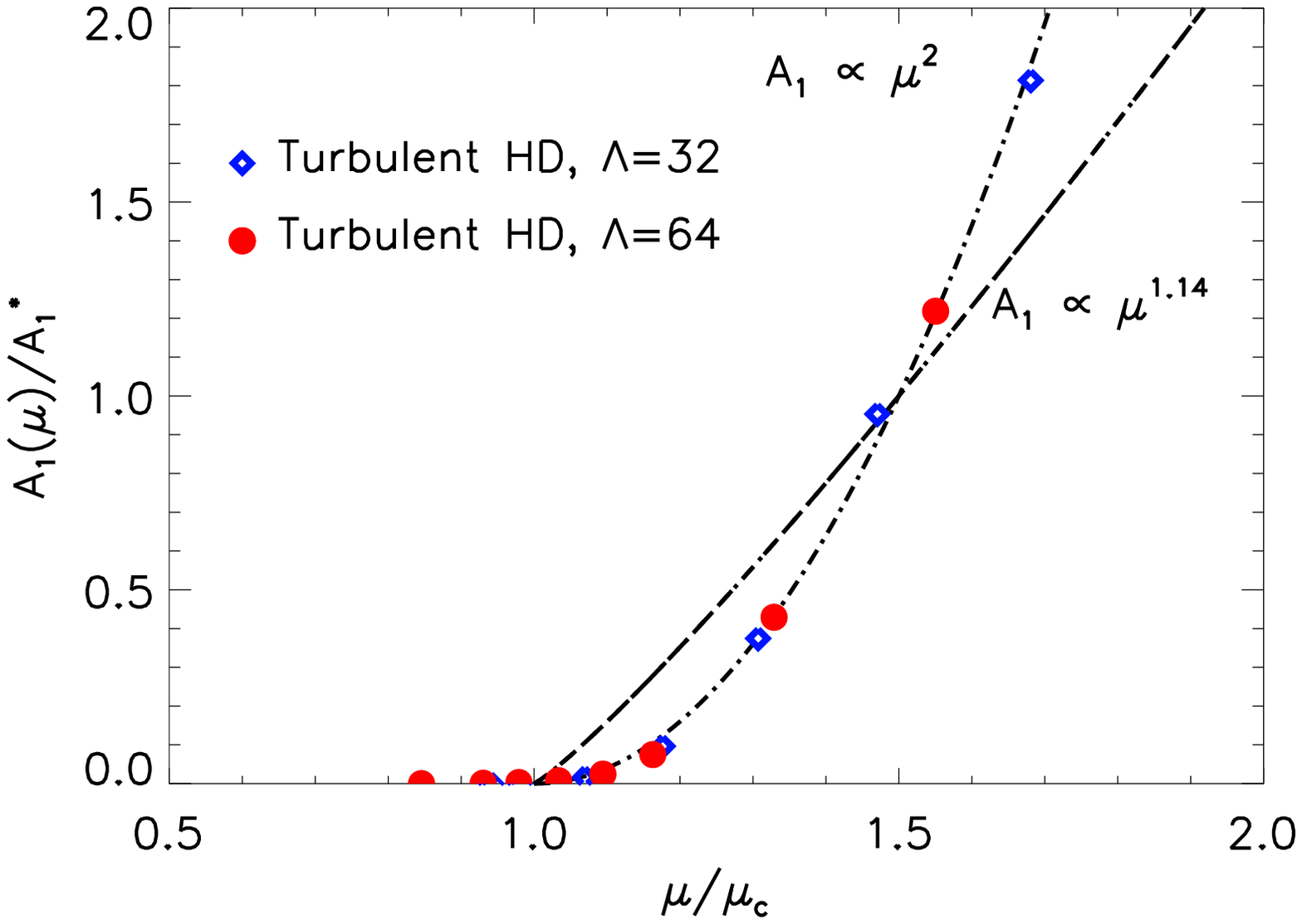}
\caption{Left panel: First moment $A_1$ for the thin layer dynamo problem for $\gamma=1$ for which
turbulent diffusion is much more effective than in the case of Fig. \ref{fig:2Dgrowth2}.
Right panel: First moment $A_1$ for the thin layer problem for the turbulent flow. }\label{fig:2Dgrowth4} 
\end{figure}                                                                                
In Fig. \ref{fig:2Dgrowth4} we show the behavior of $A_1$ for the dynamo problem as in Fig. \ref{fig:2Dgrowth2} but with a smaller value of $\gamma=1$. The anomalous exponent observed in Fig. \ref{fig:2Dgrowth2} is not present in the case of the left panel of Fig. \ref{fig:2Dgrowth4} and the data are much better fitted with the mean field exponent $\beta_1=1/2$. Similarly the second moment $A_2$ (not plotted) here is much closer to $\beta_2=1$. Finally, the energy distribution shown in panel (c) of Fig. \ref{fig:fields} does not show the  spatial distribution observed in the other panels. This observed mean-field behavior is however due to finite size effects. The anomalous scaling, and the associated intense localization of the field, are expected to be recovered in a larger system $\Lambda\to \infty$.


Furthermore, the predicted  exponents based on Eq.(\ref{eqfield}) are valid  when the noise is  white. Their values differ when the noise has different properties (see for instance \cite{halpin} p 285, and \cite{petrelis2012anomalous,alexakis2012critical}). Indeed when we simulate Eq.(\ref{eqfield}) with colored noise larger exponents are observed. The value of the measured exponents appeared to depend on the spectral properties of the noise.
This is important because in turbulent flows the spatio-temporal correlations of the fluctuations are far from being white and Gaussian. In contrast to the random flow for which the fluctuations are localized in scale, the energy cascade in the turbulent system leads to fluctuations across a wide range of scales. The exponents measured for the fully turbulent flow, such as the one depicted in the right panel of Fig. \ref{2Dflow1}, 
thus  differ from the predictions of Eq.  (\ref{eqfield}) with a white noise.  In the right panel of Fig. \ref{fig:2Dgrowth4}, $A_1$ is displayed as a function of $\mu$ for the hydrodynamic model in the turbulent state with $Re\simeq100, R_\alpha\simeq 30$ and for the same values of $\Lambda$ as in Fig. \ref{fig:2Dgrowth2}. The data overlap again in one master curve. The measured exponent is larger than both the mean field prediction and the prediction of the white noise model of Eq. (\ref{eqfield}) and is closer to $\beta_1\simeq 2$. 
Theoretical predictions for Eq. (\ref{eqfield}) in 2D with colored noise are still limited.
Understanding the precise value of these exponents from properties of the KPZ equations subject to colored spatio-temporal noise related to the spectral properties of turbulent flows would be of great interest.
  Results for KPZ in 2D are still limited but in 1D, it is known that the roughness exponent ($\chi$ with the notation of \cite{Tu})  increases with the slope $\rho$ of the noise spectrum (assumed  to be of the form $k^{-2 \rho}$). Assuming that this remain true in 2D, and using the known scaling relations  for the problem of multiplicative noise  \cite{Tu}, we expect that the exponent $\beta$ is larger when $\rho$ is large (for a turbulent flow and a noise term proportional to the velocity gradient $\rho=1/3$) than when the noise is white ($\rho=0$) which has same exponent as in the case of the random flow. Thus the exponent can be sensitive to the spatial properties of the turbulent fluctuations and in particular the existence of an inverse cascade. 

Furthermore, the universality class can depend on the vectorial or scalar form of the bifurcating field \cite{tauber2014critical}. In the examined cases it is a vector for the two physical systems. For the field equation we have observed qualitatively similar results for both a 2D vectorial and a scalar field. Further work and investigations are of course in order to clarify if there are differences in this case too that could not be resolved by the present data.

Finally we note that the considered systems are essentially 2D, and we expect that 3D systems belong to  different universality classes. Further investigations of field equations as Eq. (\ref{eqfield}) are required to determine the role of the dimension of space and of the order parameter, as well as finite size effects and long-temporal and long-spatial noise correlation effects. Further numerical but also experimental investigations are also indispensable for clarifying all aspects of this transition. We believe that the results presented in this article open new directions for the study of a variety of instabilities occurring over a turbulent system such as in turbulent atmospheric layers, surface waves driven by turbulent winds in the ocean and magnetic dynamo field generation in stars driven by turbulent convection.

\acknowledgments 
This work was granted access to the HPC resources of MesoPSL financed by the Region
Ile de France and the project Equip@Meso (reference ANR-10-EQPX-29-01) of the programme Investissements d’Avenir supervised by the Agence Nationale pour la Recherche and the HPC resources of GENCI-TGCC \& GENCI-CINES (Project No. A0070506421) where the present numerical simulations have been performed. This work has also been supported by the Agence nationale de la recherche (ANR DYSTURB project No. ANR17-CE30-0004). SJB acknowledges funding from a grant from the National Science Foundation (OCE-1459702).

\bibliography{references}
\end{document}